\documentclass[usenatbib]{mn2e}

\usepackage[psamsfonts]{amssymb}
\usepackage[dvips]{graphicx}
\usepackage{amsmath,alltt}
\def\etal{{\it et al.}}
\def\msun{M_\odot}
\def\rsun{R_\odot}
\def\D{\Delta}
\def\Dt{\D t}
\def\Dx{\D x}
\def\p{\partial}
\def\pt{\p t}

\def\bF{\boldsymbol{F}}
\def\bS{\boldsymbol{S}}
\def\bn{\boldsymbol{n}}
\def\bu{\boldsymbol{u}}
\def\bx{\boldsymbol{x}}

\title[]{A comparison of hydrodynamics techniques for modelling collisions between main sequence stars}

\author[Hy Trac, Alison Sills, and Ue-Li Pen]
{Hy Trac$^1$\thanks{Email: htrac@princeton.edu},
Alison Sills$^2$\thanks{Email: asills@mcmaster.ca}, 
and Ue-Li Pen$^3$\thanks{Email: pen@cita.utoronto.ca}\\
$^{1}$Department of Astrophysical Sciences, Princeton University, Princeton, New Jersey 08544, USA\\
$^{2}$Department of Physics and Astronomy, McMaster University, Hamilton, Ontario L8S 4M1, Canada \\
$^{3}$Canadian Institute for Theoretical Astrophysics, University of Toronto, Toronto, Ontario M5S 3H8, Canada}

\begin{document}
\maketitle

\begin{abstract}

An Eulerian TVD code and a Lagrangian SPH code are used to simulate the off-axis collision of equal-mass main 
sequence stars in order to address the question of whether stellar mergers can produce a remnant star where the 
interior has been replenished with hydrogen due to significant mixing.  Each parent main sequence star is chosen 
to be found near the turnoff, with hydrogen depleted in the core, and is modelled with a $M=0.8\msun$ realistic 
stellar model and as a $n=3$ polytrope.  An ideal fluid description with adiabatic index $\gamma=5/3$ is used for all 
hydrodynamic calculations.  We found good agreement between the simulations for the polytropic case, with the 
remnant showing strong, non-local mixing throughout.  In the interior quarter of the mass, $\sim35\%$ is mixed in 
from larger radii and on average the remnant is $\sim50\%$ fully mixed.  For the realistic model, we found less 
mixing, particularly in the interior and in the SPH simulation.  In the inner quarter, $\sim20\%$ of the contained mass 
in the TVD case, but only $\sim3\%$ in the SPH one is mixed in from outside.  The simulations give consistent 
results for the overall profile of the merger remnant and the amount of mass loss, but the differences in mixing 
suggests that the intrinsic difference between grid and particle based schemes remains a possible artifact.  We 
conclude that both the TVD and SPH schemes can be used equally well for problems that are best suited to their 
strengths and that care should be taken in interpreting results about fluid mixing.

\end{abstract}

\section{Introduction}

In dense stellar systems, such as globular clusters, galactic nuclei and star forming regions, direct collisions between stars occur quite frequently. These collisions can modify the stellar populations of the system by creating objects \citep[e.g. blue stragglers, cataclysmic variables and millisecond pulsars;][]{B95}, or by destroying
objects \citep[e.g. bright giants;][]{BD99}. These kinds of strong interactions between stars can also modify the overall evolution of the system by changing the dynamics of the system and its energy budget. The study of stellar collisions has become critical to the study of dense stellar systems \citep{Setal03}.

In order to understand how collisions can modify stellar populations
and dynamics, we need to understand the collisions themselves. We need
to understand when collisions will destroy the two stars involved, and
when a new star will be created. If a star is created, how much mass
has been lost from the system? What is the composition and structure
of the new star? The answers to these questions depend on the two
stars involved in the collision, their impact parameter, and the
velocity of the collision.

Consider for example, blue straggler stars, which are probable stellar collision products.  These are stars in clusters that are bluer, brighter, and more massive than the main sequence turnoff, and hence must be rejuvenated stars.  We see blue stragglers in all globular clusters that have been observed carefully, and they are also found in many open clusters and other environments. The number is typically a few 10's per cluster, but can get as high as a few hundred. Based on their numbers, these stars must have main sequence lifetimes on the order of Gyrs, which implies a significant amount of hydrogen in their cores. Is that hydrogen mixed into the core during the stellar collision, or is it there already? Only a detailed understand of the hydrodynamical mixing processes can answer that question.  Modelling the collision itself is a hydrodynamical problem because of the timescales involved. The merger timescale is typically on the order of hours, over which time the fluid is very close to ideal, with no radiative or diffusive effects of importance.

The first hydrodynamic simulations of stellar collisions were done
with low-resolution grid codes, in one or two dimensions
\citep{M67}. These earliest simulations were head-on collisions
between equal-mass main sequence stars at a variety of impact
velocities. Attention quickly turned to collisions between compact
objects (white dwarfs or neutron stars) and main sequence stars
\citep[e.g.][]{SS77, SS78}, and collisions involving giant
branch stars \citep{T85}. Grid methods are used today mostly for
collisions involving neutron stars, where general relativity needs to
be considered \citep[e.g.][]{CW02}.  

Smoothed particle hydrodynamics (SPH) was invented independently by
\citet{L77} and \citet{GM77}.  The first SPH simulations of stellar
collisions were published in 1987 by \citet{BH87}, who looked at a
variety of collisions involving equal-mass main sequence stars. Since
then, the SPH method has been applied extensively to a wide range of
collision scenarios. A very thorough review of collisions involving
main sequence stars can be found in \citet{FB05}. Almost all possible
combinations of stellar types have been collided, from pre-main
sequence stars \citep{LS05} through giants \citep[e.g.][]{RS91} and
even black holes and quark stars \citep{KL02}. 

The SPH method has some intrinsic advantages for studying stellar
collisions. The main advantage is that the SPH particles follow the
fluid flow. Therefore, all the computational effort is spent on the
regions of interest rather than in calculating empty grid cells. Some
stellar collisions, particularly those with large impact parameters
and low relative velocities, result in a very slow in-spiral of the
two stars, with many periastron passages and large apastron
distances. A grid code typically has to sacrifice resolution for
coverage of the large physical space traversed by the stars involved
in the collision. 

However, the SPH method also has some limitations which could affect our understanding of the outcome of the collision.  Shocks are often not well resolved, in particular the weak ones which are expected in stellar mergers.  Artificial viscosity is implemented to prevent unphysical oscillations, but it can cause spurious exchange of angular momentum between shear layers in a flow.  A careful choice of the formulation of the artificial visocity is needed to address these concerns.  In addition, there is usually no treatment of turbulent mixing.  Since the final structure and subsequent evolution of a collision product can depend quite sensitively on its composition, getting the amount of fluid mixing right is very important.

In grid codes, cells numerically force mixing of all matter that
enters. This will result in overmixing. On the other end of the scale,
in SPH codes, particles have a definite identity and entropy. Mixing
in real fluids occurs through a turbulent cascade, where at the end
some fluid elements gain entropy and some lose entropy. This process
is not permitted in SPH, except in a coarse grained sense. In the
presence of gravity, the residual identity could lead to sedimentation,
which can undo the coarse grained mixing. From a physical perspective,
we expect grid codes to overmix, and SPH to undermix, with the true
answer in between.

Some comparisons between SPH and grid codes have been done for stellar collisions \citep{Davies93}.  They collided a low-mass unevolved main sequence star (modelled as a n=3/2 polytrope) with a white dwarf. They found that the global values of the simulations were in good agreement, but that some local values (e.g. the density distribution near the white dwarf) were slightly different. They also found that the largest differences between their simulations was caused by different treatments of the potential well of the white dwarf, which was treated as a point mass in both simulations. The treatment of entropy and its effect on mixing was not investigated in that paper.

Comparisons between SPH and grid codes for other hydrodynamic problems
have also been published. For example, \citet{BB97} looked at
resolution requirements for both methods in a molecular cloud collapse
calculation \citet{DK97}. They compare both first- and second-order
hydrodynamic simulations of a collapse of a protostellar binary
system with SPH, and find that the second-order codes produce results
very similar to the SPH results. The main differences occur in the
treatment of rotation. Accretion flows were studied with both SPH and
TVD methods by \citet{MRC96}, who found that the general
characteristics of the flow were the same, although the details were
slightly different. They found that the location of the shocks were
probably better obtained in SPH, but that the shocks themselves were
better modelled using TVD. Comparisons between SPH and Eulerian codes
were also applied to dynamical bar instabilities in a rotating star by
\citet{SHC96}, and in a variety of cosmological simulations by
\citet{KOCRHEBN94}.  The overall result of these studies is that both
SPH and grid codes provide accurate descriptions of the hydrodynamics
of most astrophysical situations. However, different scenarios are
better treated with different methods, depending on the science that
one wishes to do with the simulation. 

In this paper, we compare the results of stellar collision calculations performed using an Eulerian TVD grid code \citep{TP03} and a Lagrangian SPH code \citep{Sills02}.  We have simulated the off-axis collision of two equal-mass main sequence stars with $M_0=0.8\msun$ and $R_0=0.955\rsun$.  Each parent main sequence star is chosen to be found near the turnoff, with hydrogen depleted in the core, and is modelled with a $M=0.8\msun$ realistic stellar model and as a $n=3$ polytrope.  We concentrate on a comparison of the mixing characteristics in the collision products in order to address the question of whether stellar mergers can produce a remnant star where the interior has been replenished with hydrogen due to significant mixing.  In sections 2 and 3 we describe the two hydrodynamic methods and in section 4 the numerical setup.  Section 5 gives the comparisons and we draw conclusions from those results in section 6.

\section{The Eulerian TVD Code}

A three-dimensional Eulerian hydrodynamic code \citep{TP03} is used to compute the flux of the conserved fluid 
variables of mass, momentum, and total energy across a Cartesian grid in discrete time steps.  For simplicity of 
notation, we can write the Euler equations in vector form as
\begin{equation}
\frac{\p\bu}{\pt}+\frac{\p\bF(\bu)}{\p\bx}=\bS(\bu),
\end{equation}
where $\bu$ contains the conserved fluid variables per unit volume, $\bF$ holds the hydrodynamic flux terms, and $
\bS$ represents the gravitational source terms.  The conserved cell-averaged quantities $\bu_{\bn}\equiv\bu(\bx_
{\bn})$ are defined at integer grid cell centres $\bx_{\bn}=(i,j,k)$ while the fluxes $\bF_{n+1/2}$ are defined at cell 
boundaries.  The TVD hydro code uses an operator splitting technique \citep{S68} to separately solve the 
hydrodynamic and gravitational terms.  Furthermore, the hydrodynamic step is dimensionally split such that flux 
updating is done in one dimension at a time.

In the hydrodynamic step, time integration is obtained using a second-order Runge-Kutta scheme, where a half-step 
is first performed,
\begin{equation}
\bu_i^{t+\Dt/2}=\bu_i^t-\left(\frac{\bF_{i+1/2}^t-\bF_{i-1/2}^t}{\Dx}\right)\frac{\Dt}{2}\ ,
\end{equation}
and then followed by a full-step,
\begin{equation}
\bu_i^{t+\Dt}=\bu_i^t-\left(\frac{\bF_{i+1/2}^{t+\Dt/2}-\bF_{i-1/2}^{t+\Dt/2}}{\Dx}\right)\Dt\ .
\end{equation}
The second-order accurate fluxes $\bF_{i+1/2}$ at cell boundaries are calculated using the total variation 
diminishing (TVD) condition \citep{TVD83}.  The hydrodynamic step is conservative in that flux taken out of one cell 
is added to the neighbouring cells and the conservation of mass, momentum, and energy therefore ensures the 
correct shock propagation.

To address the question of chemical mixing, the hydro code is implemented along with a particle-mesh scheme 
where tests particles of fixed mass flowing along the fluid velocity field lines can be used to track passively advected 
quantities like chemical composition.  The particle-mesh scheme is coupled to the Eulerian TVD scheme in the 
following manner.  The hydrodynamic scheme is dimensionally split and in order to synchronize the particles with 
the fluid, the particle-mesh scheme must also be dimensionally split.  Consider solving for the fluid flow in the $x$ 
direction.  While the fluid velocity at cell centers can be defined by dividing the cell-averaged momentum by the cell 
mass, a more robust measure is that of the velocity at which mass is flowing across cell boundaries.  The fluid 
velocity at cell boundaries can be calculated as
\begin{equation}
v_{x, i+1/2}^{t}=\frac{F_{i+1/2}^{t}}{(\rho_{i+1}^{t}+\rho_{n}^{t})/2}.
\end{equation}
where the fluxes $F=F(\rho)$ come from solving the mass continuity equation.  Both the half-step velocity field $v
(\bx)_{x, i+1/2}^{t}$ and full-step velocity field $v(\bx)_{x, i+1/2}^{t+\Dt/2}$ are needed for the particle-mesh scheme.  
The particles are then advected in the $x$ direction using a second-order Runge-Kutta time integration with a half 
step,
\begin{equation}
x^{t+\Dt/2}=x^t+v(\bx^t)_x^t(\Dt/2),
\end{equation}
followed by a full step,
\begin{equation}
x^{t+\Dt}=x^t+v(\bx^{t+\Dt/2})_x^{t+\Dt/2}\Dt.
\end{equation}
The velocity of the particle at position $\bx$ is interpolated from the fluid velocity field using a `cloud-in-cell' (CIC) 
technique \citep{HE88}.  The CIC interpolation is also used to determine local quantities such as density, pressure, 
entropy, and gravitational potential associated with each particle.  The coupling of the particle-mesh algorithm to the 
Eulerian TVD code has the benefit of being able to track fluid variables like in an SPH code, while still being able to 
solve the fluid equations using the Eulerian approach.

\section{The Lagrangian SPH Code}

The SPH code is the parallel version of the code described in detail in \citet{BBP95}. It was parallelized using
OpenMP. Our three dimensional code uses a tree to solve for the gravitational forces and to find the nearest 
neighbours \citep{BBCP90}. We use an adiabatic equation of state, and the thermodynamic quantities are evolved 
by following the change in internal energy using the following equation:
\begin{equation}
\frac{du_i}{dt} = \sum_{j=1}^{N} m_j \left(\frac{P_i}{\rho_i^2} + \frac{1}{2} \boldsymbol{\Pi}_{ij}\right)\boldsymbol{v}_
{ij} \cdot \nabla_i \boldsymbol{W}_{ij} ,
\end{equation}
where $u$ is the internal energy of a particle, $m$ is its mass, $P$ is pressure, $\rho$ is density, $\boldsymbol{\Pi}$ 
is the artificial viscosity, $\boldsymbol{v}$ is velocity, and $\boldsymbol{W}$ is the SPH smoothing kernel. Both the 
smoothing length and the number of neighbours can change in time and space.  We used the Balsara \citep{Bal95} 
form of artificial viscosity with $\alpha=\beta=5/6$, following the advice of \citet{LSRS99}. We ran a number of 
simulations with other forms of the artificial viscosity and with different values for the coefficients of the Balsara form. 
As expected, the differences between the implementations of the artificial viscosity were slight, except for the 
completely unphysical case of no artificial viscosity. 

Standard SPH treatments which do not properly account for the change
in smoothing length as a function of time or space can either evolve
energy or entropy properly, but not the other variable. Since our SPH
code evolves the energy equation, we are likely to have errors in the
determination of entropy and therefore may have unphysical fluid
mixing in our simulations.  However, earlier simulations
of stellar collisions done with this code (Sills et al. 2002) have
been compared to simulations by Lombardi et al. (1997) performed with
an SPH code that evolves the entropy equation. The resulting hydrogen
abundance profiles of the collision products were almost identical,
and so we feel that the difference between the entropy and energy
treatment of the thermodynamics is minimal in these circumstances.

\section{Hydrodynamic Simulations}

\begin{figure}
\begin{center}
\includegraphics[width=3.2in]{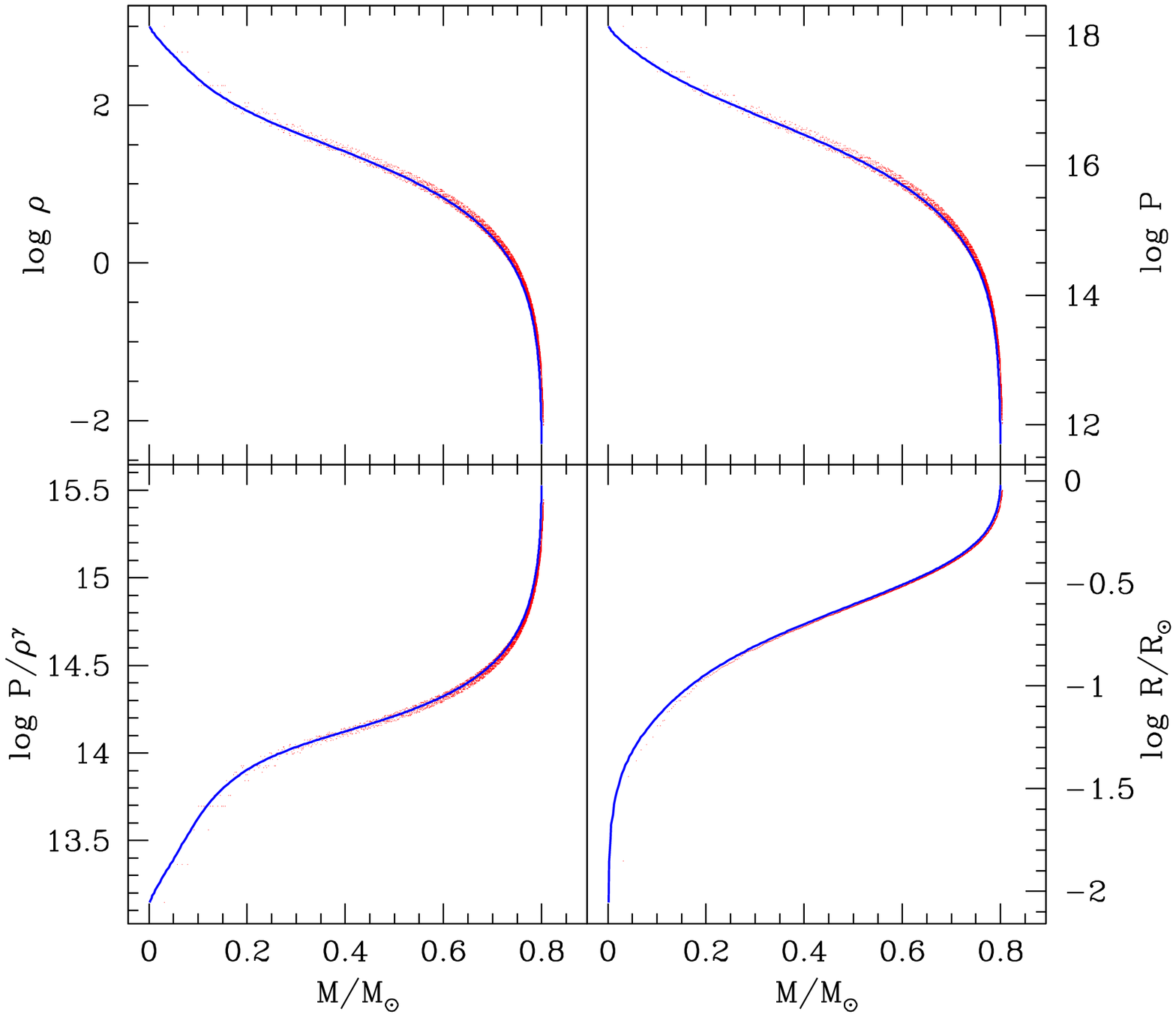}
\includegraphics[width=3.2in]{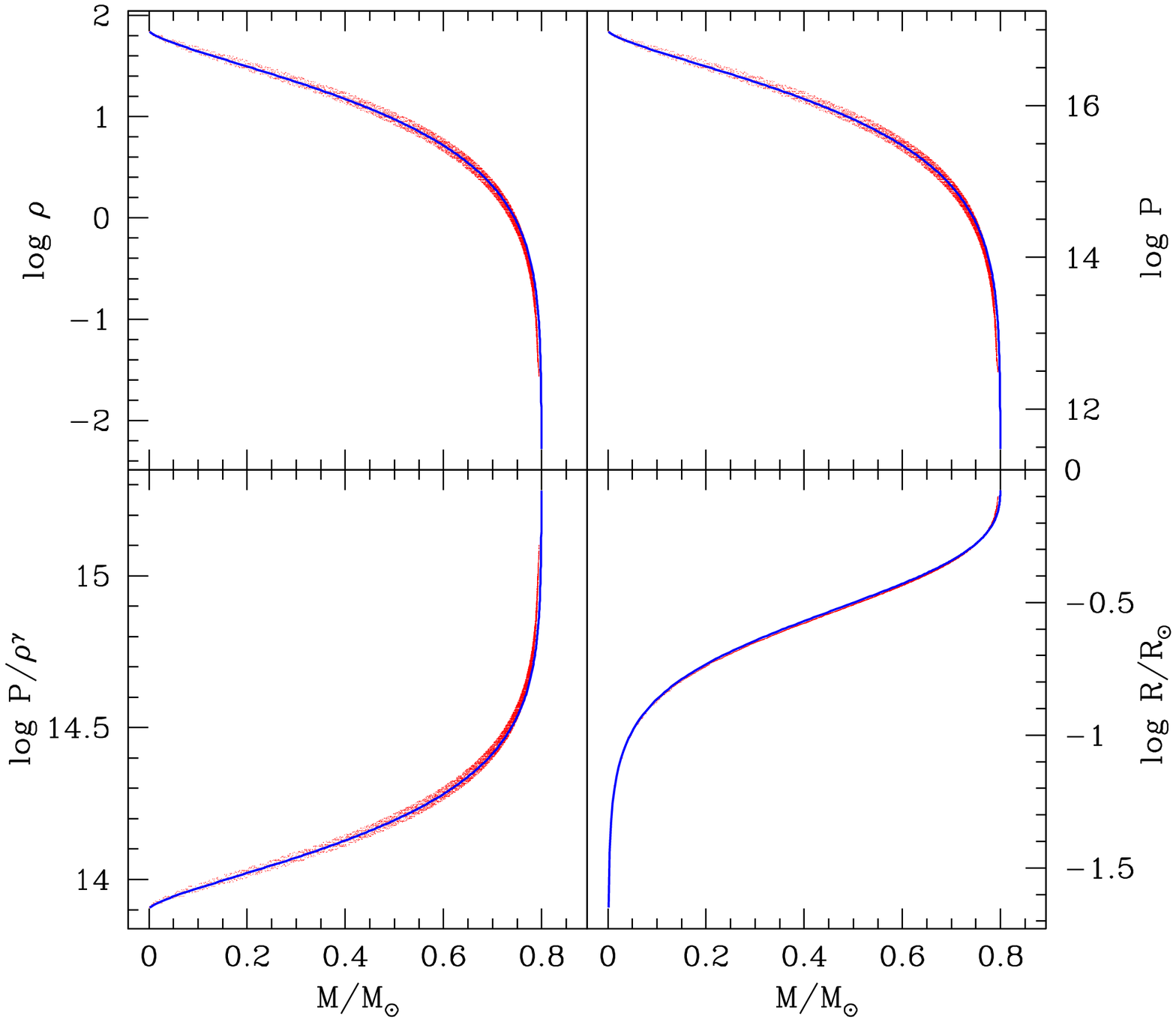}
\end{center}
\caption[]{Initial mass profiles of a $M_0=0.8\msun$ and $R_0=0.955\rsun$ main sequence star constructed with (a) a realistic stellar model (solid lines) and (b) as an $n=3$ polytrope (dashed lines).  The TVD (blue) initial conditions 
exactly match the input models and the SPH (red) initial conditions are also in very good agreement.  The SPH particles have been randomly subsampled by a factor of 5 for clarity.  The entropic variable $A\equiv P/\rho^\gamma$, where $\gamma=5/3$ is the ideal gas adiabatic index, is used in place of the entropy.  Units are in cgs.}
\label{fig:initprofile}
\end{figure}

We have simulated the off-axis collision of two equal-mass main
sequence stars with $M_0=0.8\msun$ and $R_0=0.955\rsun$.  The parent
stars are chosen to be found near the turnoff, with hydrogen depleted
in their cores, in order to address the question of whether stellar
mergers can produce a remnant star where the hydrogen core has been
replenished from significant chemical mixing.  The main sequence stars
are constructed using two different methods and the initial mass
profiles are compared in Figure \ref{fig:initprofile}.  First, a
realistic model is calculated with the Yale stellar evolution code
\citep[YREC;][]{Guenther92}.  The YREC stars are found near the
turnoff with an age of 13.5 Gyr.  In the central $\sim25\%$ of the
mass, formerly the hydrogen core, the hydrogen has been mostly
converted to helium.  Second, a polytropic model with index $n=3$
gives a good approximation to main sequence stars near the turnoff
that have relatively little mass in their convective envelopes.  Note that in
contrast, n=3/2 polytropes are good approximations of convective
regions, and very good for fully convective or mostly convective
unevolved low mass stars.  In both of our chosen stellar models, approximately 90\% of the
total mass is contained within $r<0.5R_0$.  However, the realistic
model is more centrally condensed with steeper gradients than the
polytropic one.

The parent stars are set up on zero-energy parabolic orbits with a pericentre separation equal to $0.25R_0$ and the initial trajectories are calculated assuming point masses.  The orbital plane is taken to coincide with the $x-y$ plane and the stars are initially separated by a distance of $3.75R_0$.  The dynamical time for the
stars, defined to be
\begin{equation}
\tau_{\rm dyn}\equiv\frac{1}{\sqrt{G\bar{\rho}}},
\end{equation}
where $\bar{\rho}$ is the average density, is approximately 1 physical hour and thus, modelling of the collision is a hydrodynamic problem.

In the Eulerian simulations, the collision is computed on a high
resolution Cartesian grid with $1024^3$ cells.  Initially,
each parent star has a radius of $R_0=192$ grid cells and is assigned
$N_p=256^3$ test particles.  The hydro grid has non-periodic boundaries which allow only one-way transmission 
and gas exiting the grid can not re-enter.  It is important that the parent stars are chosen to be small enough that the 
mass loss is still accurately captured on the finite grid.  The FFT Poisson solver can also be made non-periodic by 
using a gravity grid with dimensions that are twice as large as the hydro grid and treating the padded cells as
vacuum with zero density.  However, since the remnant star occupies only a fraction of the hydro grid, a more 
optimal approach is to make the Poisson solver only pseudo non-periodic and use a gravity grid with dimensions 
that are 1.5 times as large as the hydro grid.  We impose a minimum density of $10^{-6}$ g/cm$^3$ whereby gas 
cells with lower density are considered part of the vacuum and all fluxes are set to zero.  This density threshold is 
found to reduce spurious shock heating as the stars are advected across the grid \citep{TP03} and is still a few 
orders of magnitude smaller than densities of interest.  An ideal fluid approximation with an adiabatic index of $
\gamma=5/3$ is used in the simulations.  The simulations were run at NCSA \footnote{http://www.ncsa.uiuc.edu} on 
a shared-memory SGI Altix with Itanium 2 processors.  Each simulation used 64 processors and 64 GB of memory 
and required approximately 8000 cpu hours to complete.

In the SPH simulations, we used approximately 50 000 particles per
star, or 100 000 in total.  The particles were initially distributed
on an equally--spaced grid, and their masses were varied until the
density profile matched that of the stellar model we wished to
use. The smoothing length is varied to keep the number of neighbours
approximately constant at about 50. The particle masses range from
$3.2 \times 10^{-2} \msun$ to $7.8 \times 10^{-8} \msun$ with the
highest mass particles at the centre of the original star. The
simulations were run using SHARCNET \footnote{http://www.sharcnet.ca}
facilities on a 4-CPU shared-memory node of a Compaq SC40 Alpha with
667 MHz processors. Each simulation used less than 2 GB of memory and
required approximately 150 CPU hours. Note the substantially smaller
resource requirements of the SPH method for this kind of simulation
compared to the grid method.

\section{Results}

\begin{figure}
\begin{center}
\includegraphics[width=3.2in]{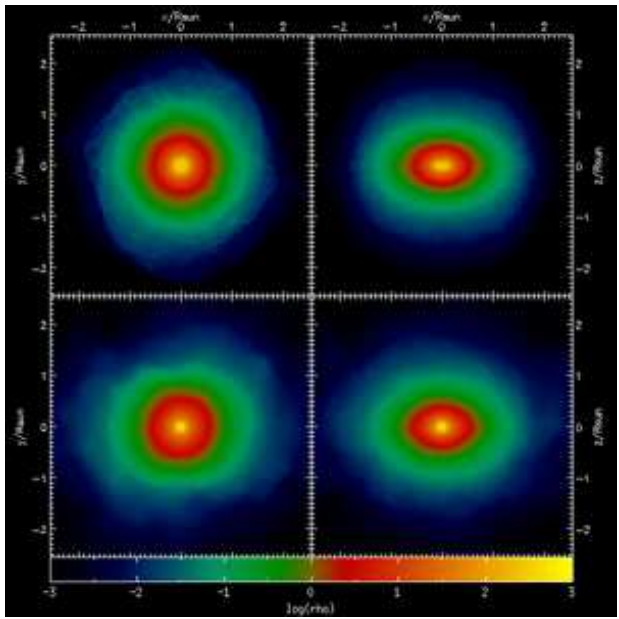}
\end{center}
\caption[]{Density plots of the merger remnants for the realistic stellar model from the TVD (top) and SPH (bottom) 
simulations.  Thin slices approximately $0.05\rsun$ thick are taken through the $x-y$ and $x-z$ mid-planes.  The orbital plane coincides with the $x-y$ mid-plane.  Both simulations produce a single rotating oblate with a mass of $M=1.5\msun$.}
\label{fig:image}
\end{figure}

Both TVD and SPH simulations of the off-axis collision between
$M_0=0.8\msun$ main sequence stars and between $n=3$ polytropes
produced a single merger remnant with mass $M_*=1.5\msun$that establishes hydrostatic
equilibrium in approximately $10\tau_{dyn}$.  Note that from here on, $M_*$ refers to the mass of bound gas and does not include ejected gas.  During the merging process, the outer envelopes of the parent stars are shock-heated and $0.1\msun$ of gas gets ejected into the interstellar medium
(ISM).  The resulting remnant is a rotating oblate that shows no
indication of having a disc.  The TVD and SPH simulations find similar mass loss with each other and also for the two stellar models.  The latter is due to the fact that both stellar models have similar structural profiles at larger radii.  We do not attempt to quantify it to higher precision because of the difficulty in precisely determining what gas is bound to the remnant.  We show some images of the remnant in Figure \ref{fig:image}, taken at time $t=10\tau_{\rm dyn}$.  All future figures show results also taken at this final time.

\subsection{Mass profiles}

\begin{figure}
\begin{center}
\includegraphics[width=3.2in]{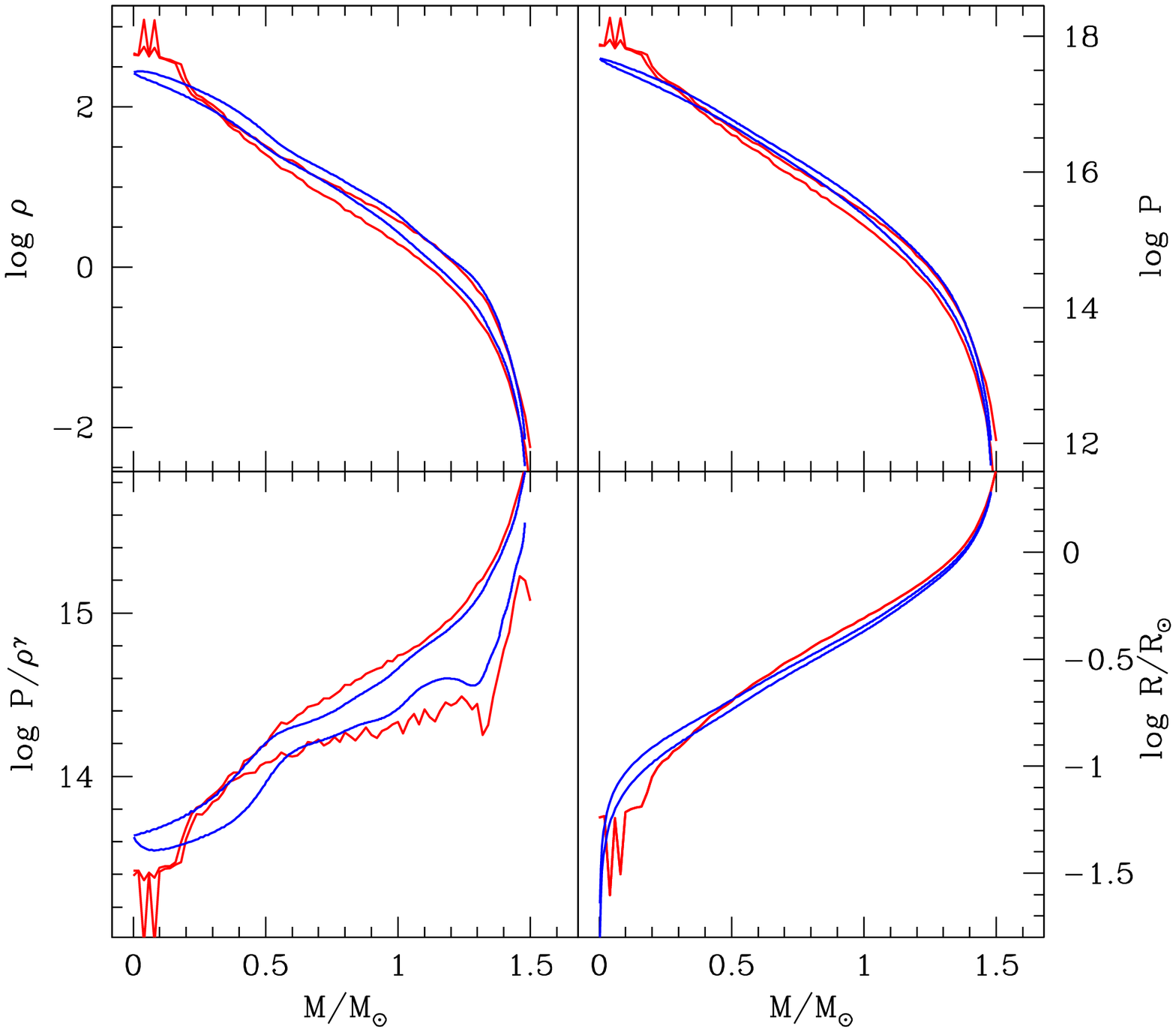}
\includegraphics[width=3.2in]{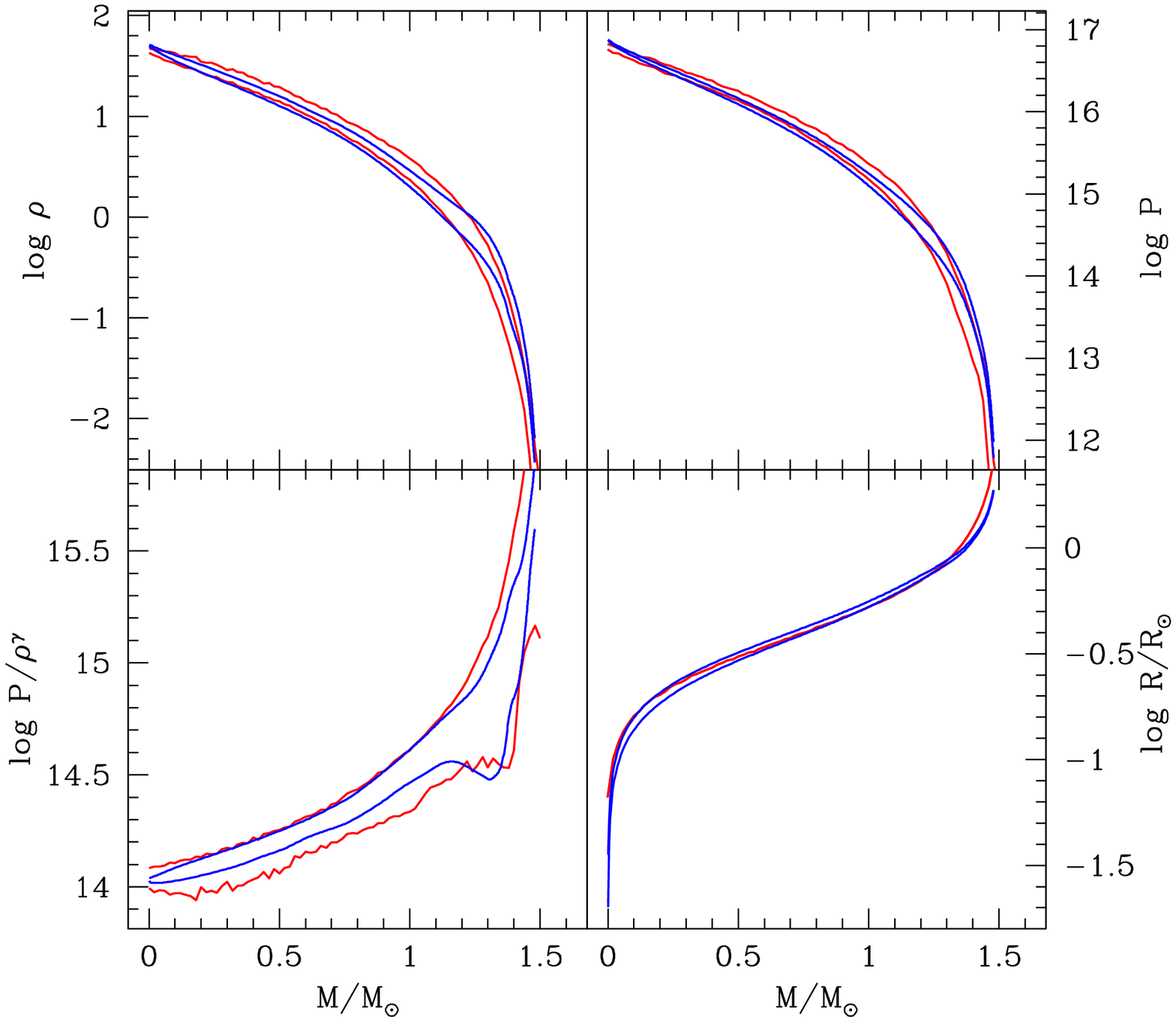}
\end{center}
\caption[]{Mass profiles of the merger remnant from the off-axis collision between equal mass $M_0=0.8\msun$ 
main sequence stars (a) and between $n=3$ polytropes (b), simulated using a TVD (blue) and a SPH (red) code.  
Note that ideal gas adiabatic index of $\gamma=5/3$ is used to define the entropic variable $A\equiv P/\rho^
\gamma$.  Units are in cgs.}
\label{fig:massprofile}
\end{figure}

Figures \ref{fig:massprofile}a and \ref{fig:massprofile}b show the mass profiles of the merger remnant from the 
collision between main sequence stars and between polytropes, respectively.  The two lines span the 1-sigma spread in the quantities in each mass bin.  In this and all future figures, the red lines give the results of the SPH simulations, while the blue lines are the TVD results. 

The profiles are constructed by first rank ordering the grid cells or
particles in terms of the effective potential and then binning the
structural variables.  The entropic variable,  
\begin{equation}
A\equiv P/\rho^\gamma ,
\end{equation}
where $\gamma=5/3$ is the ideal gas adiabatic index, is used in place of the entropy.  We also check if the particle-mesh scheme in the Eulerian code is working accurately by measuring the mass profiles for the test particles.  The CIC interpolation technique is used to assign local gas quantities to the particles and if the particles are indeed tracing the gas, then the mass profiles should be the same.  The particle-mesh algorithm does an accurate job of advecting the test particles along the velocity field lines as we find agreement at the percent level over the entire mass range.

The TVD and SPH results for the polytropic case are in good agreement at all mass scales.  The remnant star is a 
rotating oblate with a mass of $M_*=1.5\msun$ and is itself not a polytrope.  The central density and pressure have 
decreased and the entropic variable floor has risen, relative to the corresponding values in the parent polytropes. These changes can come from a combination of relaxation, shock heating, and mixing.  The initial polytropes and the final
remnant all have structural profiles that are smooth and featureless, and the modest demand on dynamic range is more than accounted for by the high resolution simulations.  Numerical relaxation in the core has been found to be only a few percent at these high resolutions.  In \citet{Trac04}, a $n=3$ polytrope with a radius of $R=96$ grid cells, which is lower resolution than that used in this paper, was advected diagonally across the $x-y$ orbital plane in over 1600 times steps to test the TVD code.  For the inner 5\% of the enclosed mass, the decrease in density and pressure and the increase in entropy was found to be $<5\%$, while for the rest of the polytrope, negligible change was found.  Similarly, a test of the SPH code, where an isolated $n=3$ polytrope was evolved for 10 dynamical times, showed that the mixing in the inner 25\% of the enclosed mass was $<2\%$.  Since shock heating is expected to be minimal in the core because of the low Mach numbers, the change in entropy suggests that some mixing may have taken place.  We will address the question of mixing in more details in section \ref{sec:massmixing}

For the main sequence stars, both codes give similar results for the
outer two-thirds of the total mass, but there are interesting
differences in the interior.  Like in the polytropic case, the remnant
star is a rotating oblate with a mass of $M_*=1.5\msun$ and some
structure is retained from that of the parent stars.  In the TVD
results, the central density and pressure are lower and the entropic variable floor is
higher compared to the corresponding SPH values.  These observed differences are
consistent with the differences between the Eulerian and Lagrangian
approaches of the two simulations.  In the Eulerian case, there will
always be some numerical diffusion between adjacent grid cells and
while this is generally alleviated by increasing the resolution, it is
difficult to fully remove this effect here.  In the hydrogen-depleted
core of the main sequence stars, the structural variables change
rapidly with mass or radius and the demand on dynamic range is
challenging for a grid code without refinement meshes.  In the SPH
results, it appears that the cores of the parent stars have settled to
the bottom of the potential well of the remnant star with little
change in entropy.  This is consistent, but not necessarily
confirmation of the argument that since the weak shocks present in
these type of mergers are not well-resolved in SPH simulations,
particles will simply follow lines of constant entropy and may
experience sedimentation.

The particular implementation of the SPH equations used in this paper
was based on non-equal-mass particles. For this reason, the inner
regions of the parent stars consist of a few massive particles, and
the density distribution requires that the particles in the inner
regions of the realistic stars are even more massive than those in the
polytrope. As a result, when the particles are binned in the collision
product, the innermost bins have very few particles, resulting in the
`teeth' seen in the top half of figure \ref{fig:massprofile}. The
resolution in the inner regions is somewhat worse than that of the
grid code simulations. We should point out, however, that the overall
results of the SPH simulations are not significantly affected by this
lack of resolution. SPH simulations that use equal-mass particles show
the same behaviour as the ones presented here, in that the cores of
the parent stars do not change their entropy and sink to the centre of
the product \citep{LRS96, LWRSW02}.
 
\subsection{Mass mixing}
\label{sec:massmixing}

\begin{figure}
\begin{center}
\includegraphics[width=3.2in]{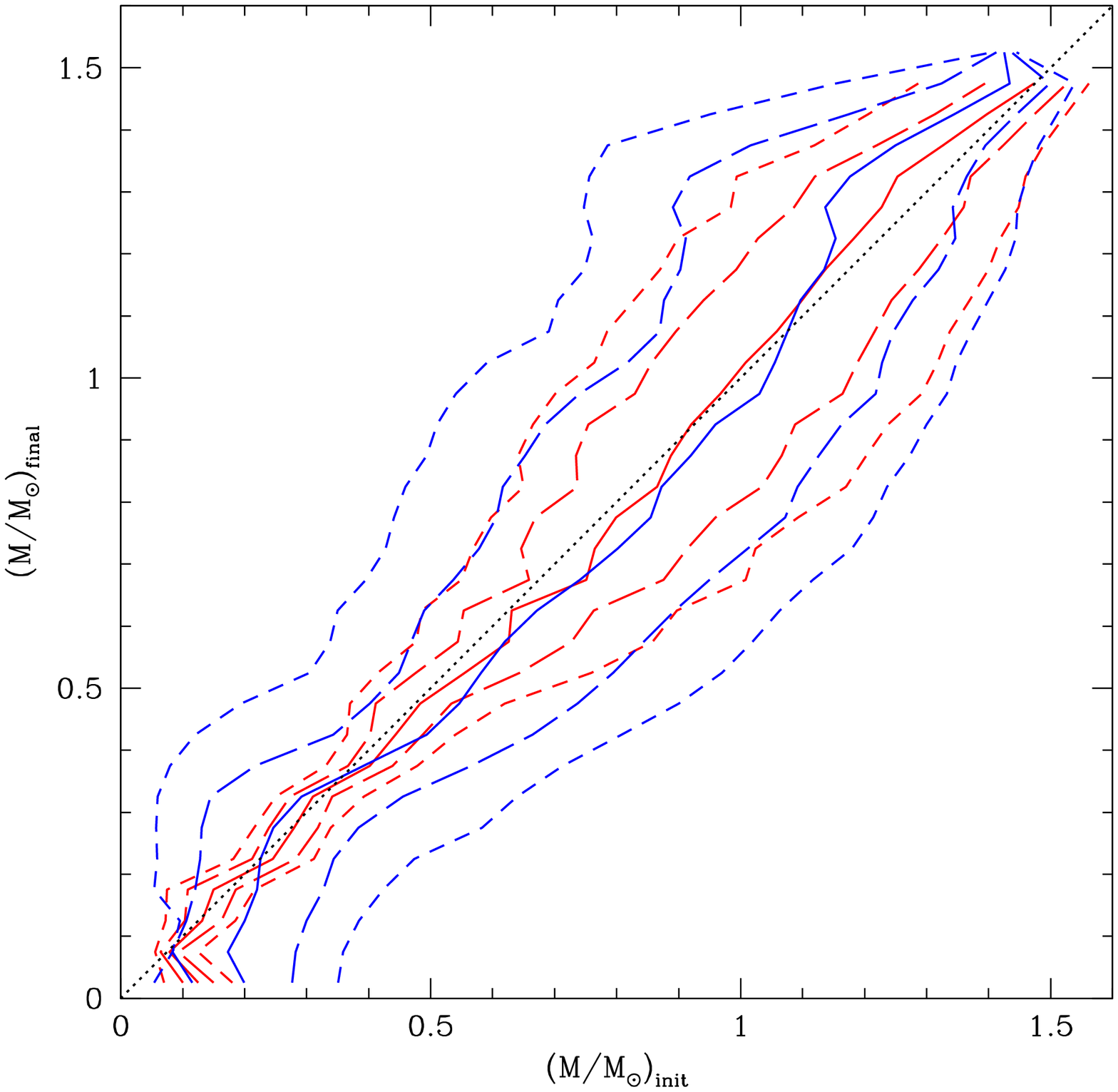}
\includegraphics[width=3.2in]{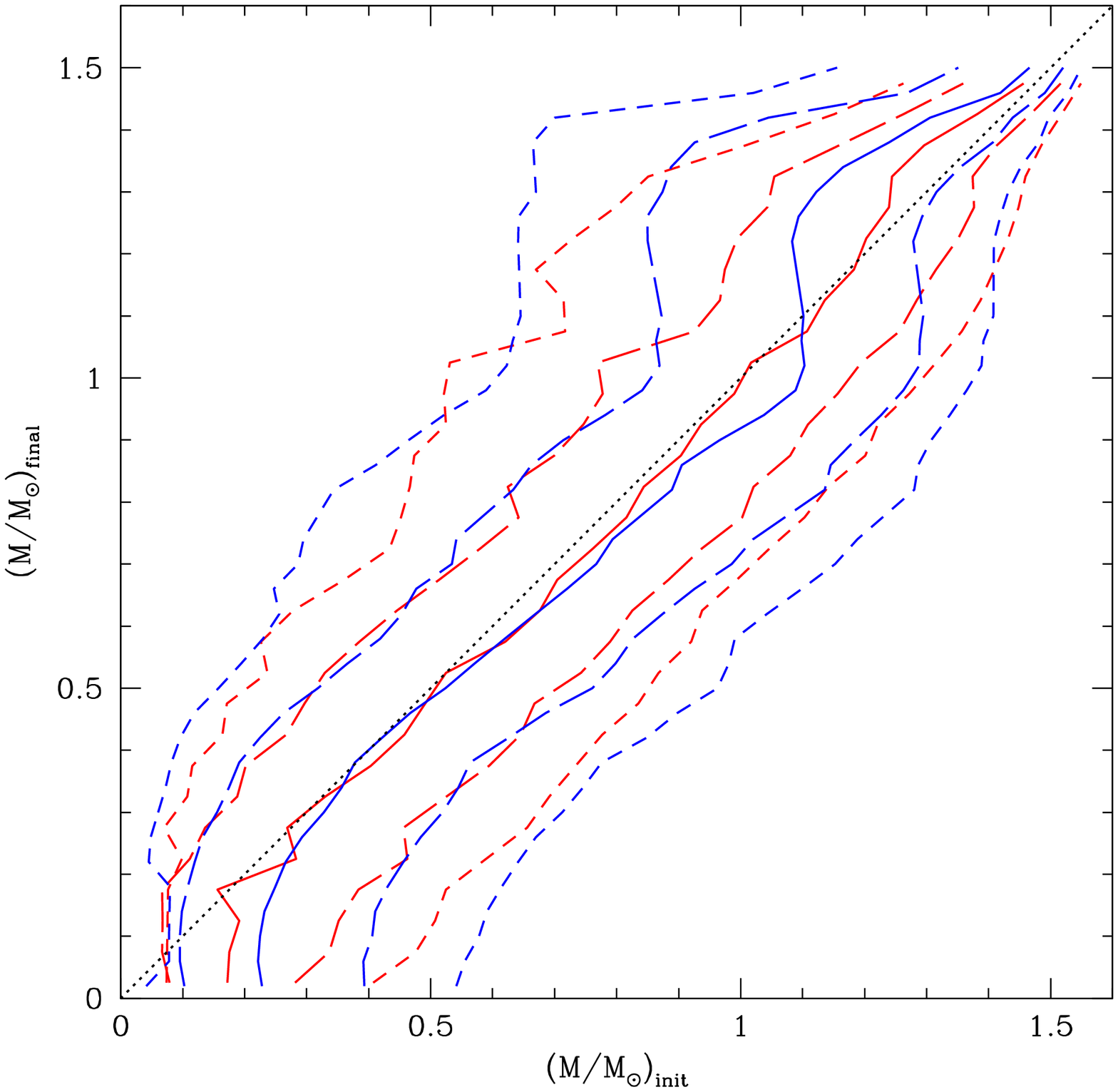}
\end{center}
\caption[]{Mass mixing found in the mergers of main sequence stars (a) and polytropes (b).  The median (solid), 50\% scatter (long-dashed), and 90\% scatter (short-dashed) values are plotted.}
\label{fig:mass}
\end{figure}

One of the advantages of Lagrangian SPH simulations is that the particles of constant mass are directly tracked 
allowing one to probe the mixing of mass and the change in entropy, and to determine the final chemical 
composition of the merger remnant.  This is now possible for Eulerian hydro codes coupled with a particle-mesh 
algorithm.  

The mixing of mass is closely examined with the following procedure.  The particles are ranked ordered using their effective potential and each particle is assigned a mass coordinate based on its ranking and this mass coordinate is simply the value of the enclosed mass.  In Figure \ref{fig:mass} the initial and final mass coordinates are compared for both the TVD test particles and the SPH particles.  The particles are binned in terms of the final mass $M_f$ in order to see where the mass in a given region of the remnant is coming from.  For each bin the 50\% value of $M_i$ is found and the median curve is illustrated by the solid line.  Similarly, the long-dashed lines enclose 50\% of the scatter, while the short-dashed lines enclose 90\% of the scatter.

Both simulations of the merger of $n=3$ polytropes produced substantial mixing.  For the inner region of the 
remnant given by $M_f/M_*<0.25$ where the replenishing of hydrogen is important, the median curve for $M_i$ 
deviates from the one-to-one line as mass is mixed into the interior, mainly from larger radii as the mixing from 
below has a hard limit.  Considering this region as a whole, 38\% of the contained mass in the TVD case and 32\% 
in the SPH one is mixed in from outside.  The average amount of mixing for the entire remnant can be estimated 
using the average width $\Delta M_i$ of the zone enclosing half of the scatter, as bounded by the long-dashed 
lines.  For complete mixing, the width is given by $\Delta M_i=M_*/2$.  In the TVD case, the remnant is estimated to 
be $\sim56\%$ fully mixed while for the SPH case, we find that it is $\sim46\%$ fully mixed.

The merger of $M_0=0.8\msun$ main sequence stars results in less mixing, particularly in the interior of the 
remnant and in the SPH simulation.  Considering the interior region $M_f/M_*<0.25$ as a whole, 21\% of the 
contained mass in the TVD case, but only 3\% in the SPH one is mixed in from outside.  On average, the TVD 
remnant is estimated to be $\sim48\%$ fully mixed while the SPH remnant is only $\sim27\%$ fully mixed.  The 
weaker mixing seen here in this case compared to the polytrope merger is due to the structural differences between 
the two stellar models.  The realistic main sequence star is more centrally condensed and it is relatively harder to 
penetrate the hydrogen-depleted core.  During the merging process, the dense interior of each parent star acts like 
a semi-compact object and remains largely intact until the two cores reach the bottom of the potential well and 
merge with one another.  The differences in mixing between the realistic and polytropic mergers have previously 
been found by \citet{SL97} for SPH simulations.  The stronger mixing seen in the Eulerian case is consistent with the 
tendency of grid codes to overmix, whereas particle codes tend to undermix.  The effect of this mixing on the 
hydrogen profile is discussed later in section \ref{sec:hydrogenmixing}.

\citet{LRS96} have also studied the mixing of mass in SPH simulations
of stellar collisions for various stellar models.  For comparison,
their closest model to ours involves the collision between $n=3$
polytropes with $M_0=1\msun$ and $R_0=1\rsun$.  When considering the
interior quarter of the enclosed initial mass $M_i/M_0$, they found a
mixing value of $\sim12\%$. Unlike our simulations, they used
equal-mass particles, and so had approximately 7000 particles in the
inner 25\% of the star. In our simulation, we have just over 1500
particles in that region of the merger remnant, despite having more
particles in our simulations overall. \citet{Sills02} show that lower
resolution simulations result in more mixing (see their figure 1),
simply because the regions of interest are by necessity more smoothed.
We conclude that our results are in good general agreement with other
SPH simulations of stellar collisions.

\subsection{Entropy change}
\label{sec:entropychange}

\begin{figure}
\begin{center}
\includegraphics[width=3.2in]{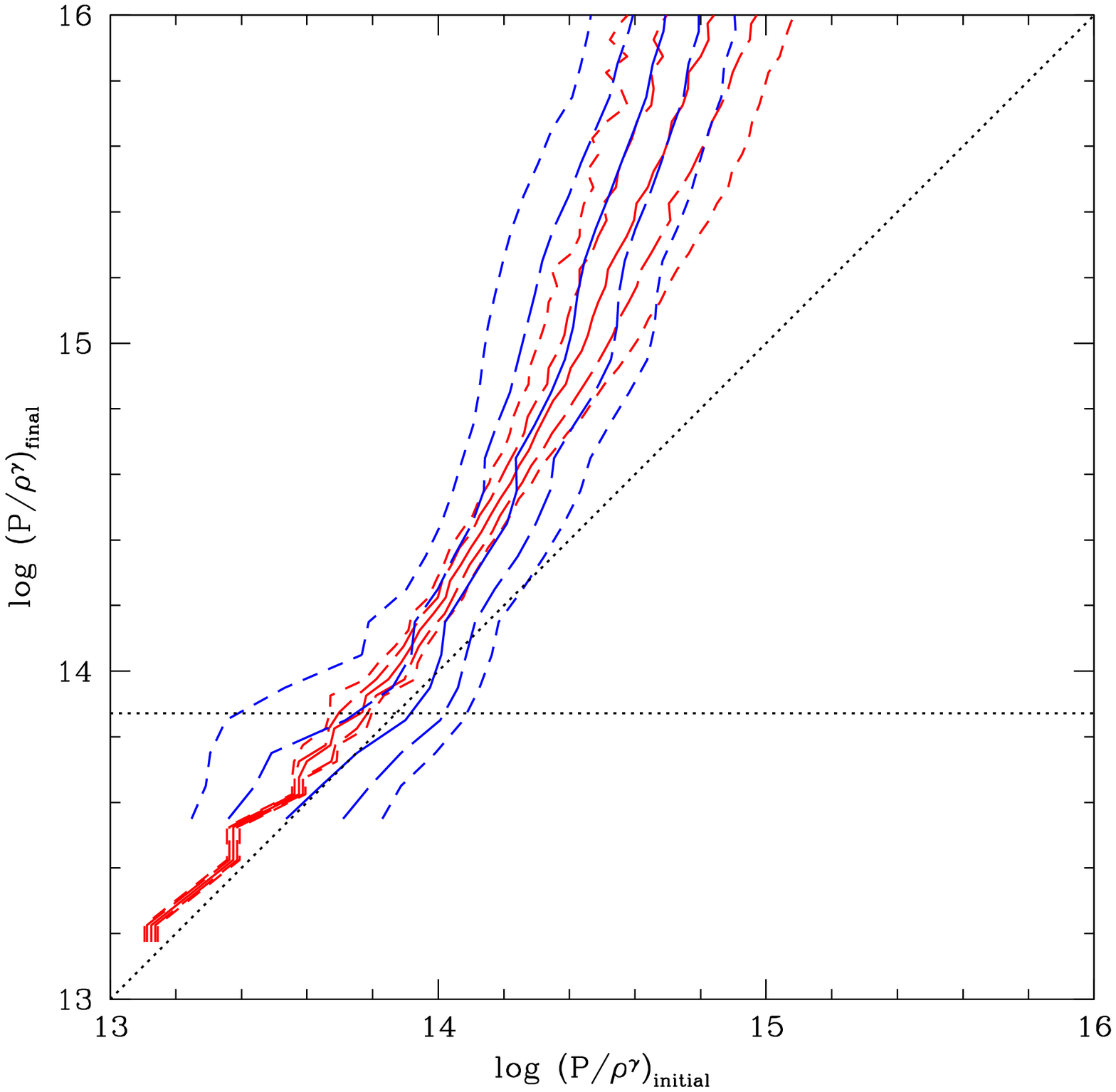}
\includegraphics[width=3.2in]{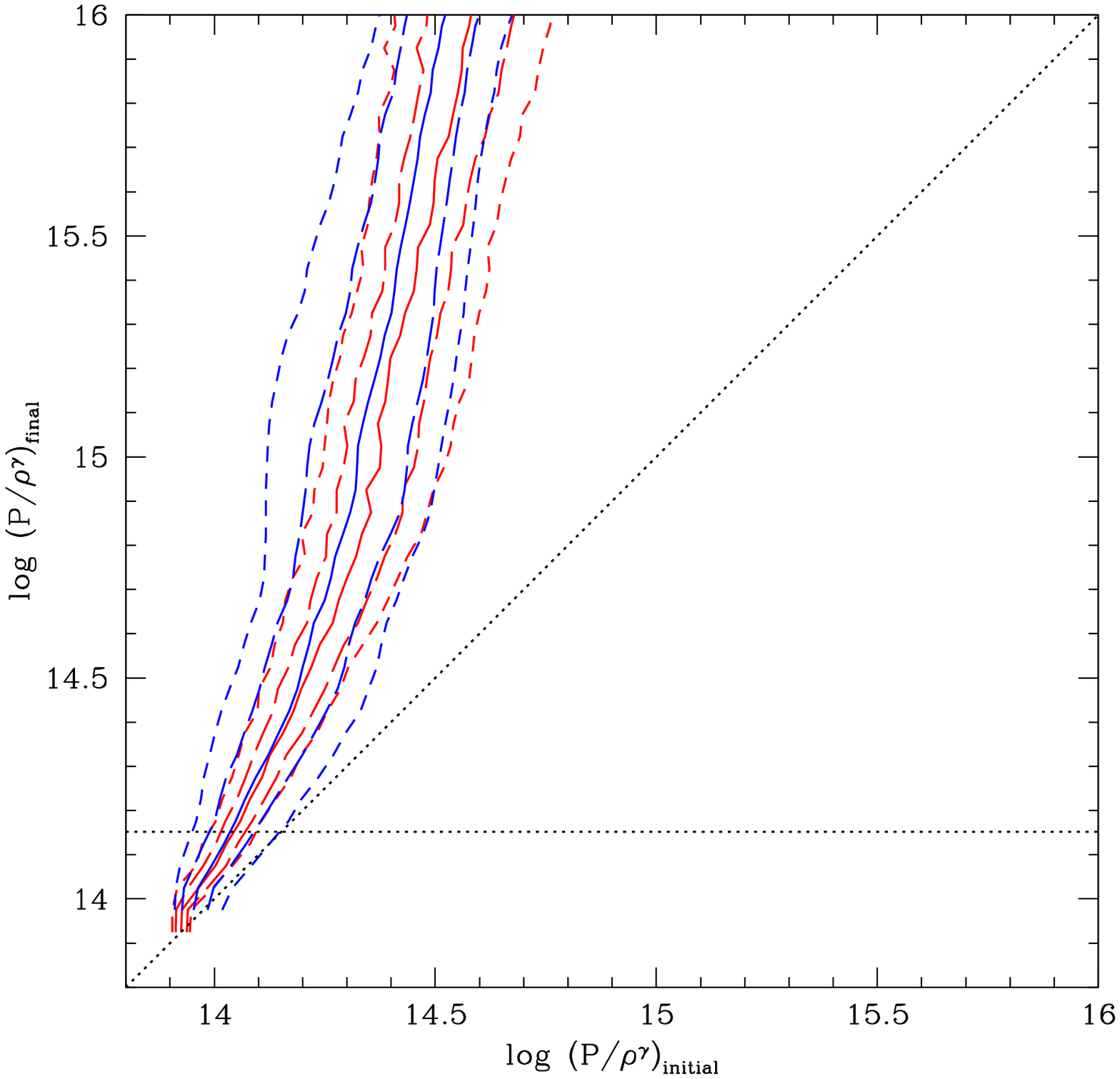}
\end{center}
\caption[]{Change in the entropic variable found in the mergers of main sequence stars (a) and polytropes (b).  The median (solid), 
50\% scatter (long-dashed), and 90\% scatter (short-dashed) values are plotted.   The diagonal dotted line 
represents no change while the horizontal dotted line gives the entropic variable for the interior quarter of the 
remnant mass $M_f/M_*=0.25$.  Note that ideal gas adiabatic index of $\gamma=5/3$ is used to define the entropic 
variable $A\equiv P/\rho^\gamma$.  Units are in cgs.}
\label{fig:entropy}
\end{figure}

In Figure \ref{fig:entropy} the initial and final values of the entropic variable $A\equiv P/\rho^\gamma$ of the TVD test particles and the SPH particles are compared.  Note that ideal gas adiabatic index of $\gamma=5/3$ is used to define the entropic variable.  In the absence of shocks, mixing, and relaxation, the fluid elements in the parent stars should be simply sorted by entropy in the remnant star, with $A_f=A_i$.  

In the polytropic merger, it is not surprising that significant change in the entropic variable is found considering the amount of mixing that has occurred.  The two simulations are also in good agreement, though the small differences suggest that the TVD remnant is more strongly mixed, as seen in the previous section, and more strongly shock heated, which is expected because grid codes are known to provide better capturing of shocks.  Note that while strong mixing has been found to occur in the interior of the remnant, it has only raised the entropy floor by a small amount.  

With main sequence stars, the TVD results show an artificial increase in the entropy floor that is caused by 
numerical relaxation rather than mixing or shock heating.  In comparison, the lowest entropy SPH particles show 
only relatively small changes with small dispersion.  Overall, there is less change in the entropic variable in the realistic merger, but it is difficult to tell from the plots how much of the entropy increase is due to mixing and how much to shock heating.  To account for the differences between Figure \ref{fig:entropy}a and Figure \ref{fig:entropy}b, mixing must play a substantial role in changing the entropy profile or there must be significantly more shock heating in the polytropic merger.  Note that for some TVD test particles where shocks have not significantly raised the entropic variable, the decrease in local entropy due to mixing is observable.

\subsection{Hydrogen profile}
\label{sec:hydrogenmixing}

\begin{figure}
\includegraphics[width=3.2in]{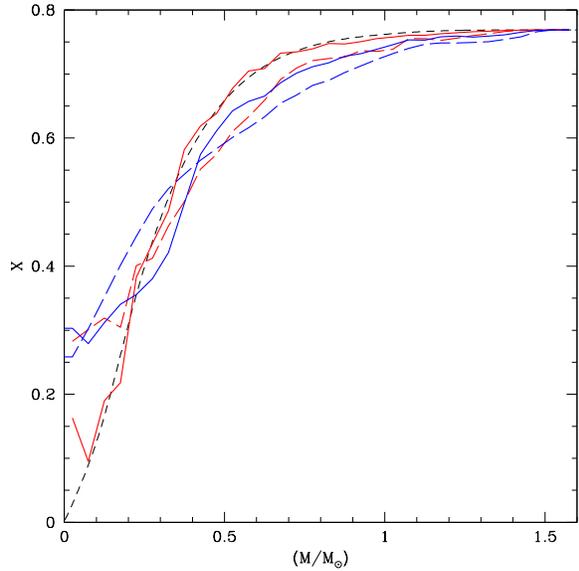}
\caption[]{Hydrogen mass fraction in the merger remnant from the collision between main sequence stars (solid) 
and between polytropes (long-dashed).  The short-dashed black line represents the hydrogen profile expected in 
the absence of mixing.}
\label{fig:hydrogen}
\end{figure}

The hydrogen mass fraction $X$ in the merger remnant from the collision between main sequence stars is plotted 
as a function of the enclosed mass in Figure \ref{fig:hydrogen}.  For comparison, the short-dashed line represents 
the hydrogen profile expected in the absence of mixing.  The SPH remnant has a hydrogen profile that closely 
resembles the no mixing case.  The hydrogen profile in the hydrogen-depleted interior is sensitive to mixing, but 
only weak mixing and small changes of $<5\%$ to the hydrogen fraction are found.  While stronger mixing is found 
at larger radii, it occurs between regions with approximately similar initial hydrogen abundance and this results in 
only small changes.  In the TVD remnant the stronger, non-local mixing has noticeably changed the hydrogen 
profile.  Considering the interior region $M_f/M_*<0.25$ as a whole, the hydrogen fraction has on average 
increased by 24\% and this is consistent with the previous finding where 21\% of the mass is mixed in from the 
outside.

We also include the results for the polytropic collision to highlight the difference in mixing, but since this is not a 
realistic model, the initial chemical composition of the polytropes was taken from that of the $M_0=0.8\msun$ main 
sequence star.  In this case, both simulations show stronger, non-local mixing and this has resulted in more 
changes to the hydrogen mass fraction throughout.

\section{Discussion and Conclusions}

The biggest difference we see between the SPH and TVD simulations is
in the amount of mixing in the interior of the remnant star.  In order
for a collision product to be observed as a blue straggler, it must
remain as a main sequence star for a reasonably long time.  The SPH
simulations predict a very small amount of mixing, and we know from
stellar evolutionary models of these collision products
\citep{SLBDRS97} that the lifetimes of these collision products are
quite short ($\sim 10^6$years). The TVD simulation of the same
collision produces a collision product which has a central hydrogen
abundance of $X\sim0.3$, which will have a significantly longer
lifetime ($\sim 10^8$ years).  It is not clear from this work which
answer is more correct. We probably need to turn to the observations
of blue stragglers in a variety of clusters. We know already that
simple blue straggler population synthesis models (based on results
from SPH simulations) have some trouble reproducing the bright end of
the observed distribution in some clusters \citep{SB99}, but we also
know that Eulerian schemes contain some numerical diffusion and hence
over-predict the mixing. The true result is probably somewhere between
these two cases.

Note that the mixing effect will be strongest in the collision that we
have chosen to model: a collision between two stars at the main
sequence turnoff. This is, of course, why we chose to compare this
particular collision, but we should caution the reader not to apply
these results uniformly to all collisions. Stars that are less evolved
and that have flatter composition profiles will show the effect of any
mixing less when they collide, since any material that is `mixed' will
have the same composition. The particular choice of hydrodynamical
method should not produce such a different result as it does for the
highly evolved case shown here.

The differences in mixing seen in the merger of main sequence stars
and the merger of polytropes are due to the structural differences
between the two stellar models.  The realistic main sequence star is
more centrally condensed and it is relatively harder to penetrate the
hydrogen-depleted interior.  Therefore, one expects that the larger
the density contrast in the parent star, the weaker the mixing for a
fixed collision velocity and geometry.  Given the differences in
results, it is important to use realistic stellar models rather than
approximations like polytropes when simulating stellar collisions
\citep{SL97}.

The Eulerian simulations have the advantage of not requiring any
artificial viscosity or the need to adjust any smoothing lengths and
therefore, the numerical solutions to the fluid equations, in
particular the capturing of shocks, are generally robust.  However,
they have the disadvantage of limited dynamic range due to the fact
that the fluid equations were solved on a single Cartesian grid.  One
consequence of this is that only a small collisional parameter space
can be considered and another is that the simulations often suffer
some level of numerical diffusion in regions with large density
gradients.  In principle, the problems can be rectified by running
larger volume and higher resolution simulations, but in practice this
is not feasible because of limited computational resources.  One
remedy is to use adaptive mesh refinement (AMR) schemes in codes like
ART \citep{ART}, Enzo \citep{Enzo}, and FLASH
\citep{Flash}.  One caution we make is that it is important to
understand any systematic problems in the transport of fluid across
refinement mesh boundaries when studying mixing during stellar
mergers.

The SPH method is the hydrodynamical method of choice for
modelling stellar collisions, because of its adaptability to a wide
range of impact parameters, velocities, stellar masses, etc. One does
not need to know where the material will go before starting the
simulation, since the SPH particles are free to travel wherever they
would like in space. The method has its drawbacks in terms of
treatment of shocks and spurious diffusion of particles caused by the
numerical methods.  In addition, turbulent mixing is generally inhibited by the addition of artificial viscosity \citep{Dolag05}.  We have shown in this paper that for the most part,
the two methods produce the same results. The structure of the
resulting collision product is very similar: the density and pressure
profiles of the models are almost identical. However, the amount of
fluid mixing is somewhat different between the two methods, which can
result in differing stellar lifetimes for the collision product. 

We conclude that both Eulerian and Lagrangian implementations of fluid
dynamics can be used to study stellar collisions, and that each code
should be applied in a way to maximize its strengths. Care should be
taken when interpreting results related to fluid mixing, but all other
properties of the system are determined correctly by both approaches.

\section{Acknowledgments}
We thank the anonymous referee for the constructive critique of the paper. A.S. is supported by NSERC.  H.T. acknowledges NASA grant NNG04GC50G. This work was made possible by the facilities of the Shared Hierarchical Academic Research Computing Network (SHARCNET: www.sharcnet.ca) and the National Center for Supercomputing Applications (NCSA: www.ncsa.uiuc.edu).

\end{document}